\pdfoutput=1
\documentclass[twocolumn,superscriptaddress,aps,preprintnumbers,amsmath,amssymb,prd,nofootinbib]{revtex4-2}

\usepackage{amsmath}
\usepackage{amssymb}
\usepackage{amsfonts}
\usepackage{graphicx}
\usepackage{xcolor}
\usepackage{xfrac}
\usepackage{comment}
\usepackage{pifont}
\usepackage{physics}
\usepackage{fourier}
\usepackage{hyperref}
\usepackage{bm}
\usepackage{enumitem}

\definecolor{rossoferrari}{HTML}{D9073D}
\definecolor{mediumblue}{HTML}{0000CD}
\definecolor{forestgreen}{HTML}{228B22}
\definecolor{desy_blue}{HTML}{009EE2}
\definecolor{desy_orange}{HTML}{FD8800}
\definecolor{light_pink}{rgb}{1,0.4,0.4}
\definecolor{light_blue}{rgb}{0.284602,0.317763,0.963947}
\hypersetup{setpagesize=false,bookmarksnumbered=true,bookmarksopen=true, colorlinks=true,linkcolor=light_blue,urlcolor=rossoferrari,citecolor=rossoferrari,linktocpage=false}

\usepackage{slashed}

\renewcommand{\thefootnote}{\fnsymbol{footnote}}

\newcommand{\bea}{\begin{array}}
\newcommand{\eea}{\end{array}}
\newcommand{\beq}{\begin{eqnarray}}
\newcommand{\eeq}{\end{eqnarray}}

\newcommand{\meV}{ \ {\rm meV} }
\newcommand{\eV}{ \ {\rm eV} }
\newcommand{\KeV}{ \ {\rm keV} }

\newcommand{\GeV}{\  {\rm GeV} }
\newcommand{\TeV}{\  {\rm TeV} }

\newcommand{\lmk}{\left(}  
\newcommand{\rmk}{\right)}
\newcommand{\lkk}{\left[}  
\newcommand{\rkk}{\right]}

\newcommand{\del}{\partial}  
\newcommand{\la}{\left\langle} 
\newcommand{\ra}{\right\rangle}

\newcommand{\Mpl}{M_{\rm Pl}}

\newcommand{\Kahler}{K\"{a}hler }
\newcommand{\cphi}{\varphi}

\def\eq#1{Eq.~(\ref{#1})}

\definecolor{orange}{RGB}{255,100,0}
\definecolor{rosepink}{RGB}{248,100,100}

\begin{document}

\title{
Axion cogenesis without isocurvature perturbations
}

\author{Raymond T. Co}
\email{rco@iu.edu}
\affiliation{Physics Department, Indiana University, Bloomington, IN, 47405, USA}
\affiliation{William I. Fine Theoretical Physics Institute, School of Physics and Astronomy, University of Minnesota, Minneapolis, MN 55455, USA}

\author{Masaki Yamada}
\email{m.yamada@tohoku.ac.jp}
\affiliation{Department of Physics, Tohoku University, Sendai, Miyagi 980-8578, Japan}
\affiliation{FRIS, Tohoku University, Sendai, Miyagi 980-8578, Japan}

\preprint{UMN-TH-4306/23, FTPI-MINN-23-26, TU-1219}

\date{\today}


\begin{abstract}
\noindent
Axion rotations can simultaneously explain the dark matter abundance and the baryon asymmetry of the Universe by kinetic misalignment and axiogenesis. We consider a scenario in which the Peccei-Quinn symmetry breaking field is as large as the Planck scale during inflation and the axion rotation is initiated by the inflaton-induced potential immediately after the end of inflation. This is a realization of the cogenesis scenario that is free of problems with domain walls and isocurvature perturbations thanks to large explicit Peccei-Quinn symmetry breaking at the Planck scale during inflation. The baryon asymmetry can be more efficiently produced by lepto-axiogenesis, in which case the axion mass is predicted to be larger than $\mathcal{O}(0.1) \meV$. 
We also discuss a UV complete model in supersymmetric theories. 
\end{abstract}

\maketitle

\renewcommand{\thefootnote}{\arabic{footnote}}
\setcounter{footnote}{0}

\section{Introduction}

The QCD axion is predicted by the Peccei-Quinn (PQ) mechanism that addresses the strong CP problem~\cite{Peccei:1977hh, Peccei:1977ur,Weinberg:1977ma,Wilczek:1977pj}. The axion has a rich phenomenology in cosmology as well as particle physics. 
It is a good candidate for dark matter (DM) because of the small couplings to the Standard Model (SM) particles and viable cosmological origins to explain the observed DM abundance. 
In conventional scenarios, 
there are two different mechanisms for the production of axions in the early universe. The first one is the misalignment mechanism~\cite{Preskill:1982cy,Abbott:1982af,Dine:1982ah}, where the coherent oscillation of the axion, namely the phase direction of the PQ-symmetry breaking field, is induced at the QCD phase transition. This can be a dominant source of axions in the case where the PQ symmetry is spontaneously broken during inflation.
The second one is the stochastic production from cosmic strings and domain walls that form after spontaneous symmetry breaking of the PQ symmetry and the QCD phase transition~\cite{Vilenkin:1984ib,Davis:1986xc}. This occurs if the PQ symmetry is spontaneously broken after inflation. A fraction of the axion density is also produced by the misalignment mechanism in this case. 
In both scenarios, the PQ-symmetry breaking scale or the axion decay constant should be of order $10^{12} \GeV$ to explain the DM abundance (see Ref.~\cite{DiLuzio:2020wdo} for a review). A higher value is possible for pre-inflationary PQ breaking if the initial misalignment angle is fine-tuned to be small. 
However, a smaller axion decay constant would be difficult to realize in the standard cosmological scenario in a simple setup.

A new production mechanism called kinetic misalignment~\cite{Co:2019jts,Chang:2019tvx} is recently introduced to explain the observed dark matter abundance for a smaller axion decay constant. 
A PQ-symmetry breaking field is assumed to have a large vacuum expectation value (VEV) during and after inflation, e.g., by a tachyonic effective mass of order the Hubble parameter. 
Once the Hubble parameter decreases to the bare mass of the PQ breaking field, the field starts to oscillate around the origin of the potential. 
At the onset of oscillations, the field is kicked in the phase direction by a higher-dimensional PQ breaking term. This dynamics produces a PQ charge, similar to the Affleck-Dine mechanism~\cite{Affleck:1984fy,Dine:1995kz}. 
The produced PQ charge is approximately conserved until the QCD phase transition, at which the charge density is converted to the number density of the axion. It was discussed that DM abundance can be explained by this mechanism for a much smaller axion decay constant than $10^{12} \GeV$. 
Moreover, the nonzero PQ charge before the QCD phase transition provides an interesting possibility for baryogenesis. The PQ charge biases the other asymmetries in the SM via transport equations. This provides a nonzero $B+L$ or $B-L$ asymmetry by the electroweak sphaleron or $B-L$-violating Weinberg operator. These scenarios are called axiogenesis~\cite{Co:2019wyp,Co:2020xlh} and lepto-axiogenesis~\cite{Domcke:2020kcp,Co:2020jtv,Kawamura:2021xpu}.

However, one must pay particular attention to the quantum fluctuations of the PQ-symmetry breaking field during inflation. The axion field value at late times is quite sensitive to the initial phase and radius of the PQ field because of the different strengths of the kick by the higher-dimensional operator. 
Even tiny initial perturbations $\delta \theta_i$ amplify over the course of the cosmological evolution, i.e.~$\delta \theta_f = \delta\dot\theta \Delta t = \dot\theta \delta \theta_i \Delta t \gg \delta \theta_i$, and may result in the domain wall problem~\cite{Co:2022qpr}. 
The small initial perturbations arise as quantum fluctuations during inflation if the mass of the phase/radial direction of the PQ breaking field is smaller than the Hubble parameter. These fluctuations can lead to the dangerous domain wall problem and/or excessive isocurvature perturbations in dark matter and the baryons as elaborated in Refs.~\cite{Co:2020dya,Co:2020jtv,Co:2021qgl,Co:2022qpr} in the context of axion rotations. The resultant isocurvature perturbation is given by $\mathcal{P}_S = \left\langle \left( \delta Y_\theta / Y_\theta \right)^2 \right\rangle$ where $Y_\theta$ is the PQ charge yield. The PQ charge quantum fluctuations $\delta Y_\theta$ can result from the fluctuations of the angular $\theta_i$ or radial field value $S_i$ given by $H_I/2\pi$ with $H_I$ the Hubble scale during inflation. Parametrically, $\mathcal{P}_S \propto H_I^2/S_i^2$ and the cosmological bound of $\mathcal{P}_S \lesssim 10^{-10}$~\cite{Planck:2018vyg} significantly constrain the scale of inflation.

In this paper, we address these issues by introducing an effective mass for the phase direction that is of order or larger than the Hubble parameter during inflation. We consider the case in which the VEV of the PQ breaking field during inflation is as large as the Planck scale. Since we expect that any global symmetries are explicitly broken by gravity~\cite{Perry:1978fd,Hawking:1979hw,Giddings:1988cx,Gilbert:1989nq,Kallosh:1995hi,Harlow:2018jwu,Harlow:2018tng}, the explicit breaking of the PQ symmetry should become effective at such a large VEV. This means that the phase direction acquires a large mass and does not fluctuate during inflation.%
\footnote{
This mechanism, providing a very large mass for the saxion and the axion during inflation, may be relevant to generate cosmological collider signals~\cite{Lu:2021gso,Chen:2023txq}. 
}
This is naturally realized in supergravity models, where Planck-suppressed higher-dimensional terms in the \Kahler potential lead to a Hubble-induced mass to the phase of the PQ breaking field~\cite{Dine:1995kz}.

This scenario works in general, but we especially consider a scenario in which the PQ breaking field starts to oscillate just after the end of inflation. 
This is realized in the case where the PQ breaking field acquires Hubble-induced mass terms via different operators during and after inflation. For example, if both the potential term and the kinetic term for the inflaton couple to the PQ breaking field, 
the effective potential of the PQ breaking field changes during and after inflation. 
This is specifically realized in supergravity models, even naturally, where Planck-suppressed terms in the \Kahler potential lead to couplings of the PQ breaking field to the potential as well as the kinetic term of the inflaton. 
Then one can consider the case with a negative Hubble-induced mass term during inflation and a positive one after inflation~\cite{Yamada:2015xyr}.%
\footnote{
See also Refs.~\cite{Kamada:2014qja,Kamada:2015iga,Hasegawa:2017jtk,Hasegawa:2018yuy} for the opposite case, where the Hubble-induced mass term is positive during inflation and negative after inflation.}
In this case, the PQ breaking field starts to oscillate at the end of inflation.%
\footnote{
If the PQ breaking field is identified as the inflaton, it apparently begins to oscillate after inflation (see, e.g., Ref.~\cite{Lee:2023dtw} in the context of the kinetic misalignment mechanism). 
In our case, however, the PQ breaking field is different from the inflaton and they start to oscillate simultaneously after inflation due to the flipped Hubble-induced masses. 
}
The dynamics of the PQ breaking field is therefore determined by the coupling to the inflaton, namely, almost independent of the parameters in the low-energy sector. 
The resulting abundance of the axion is therefore determined solely by the energy scale of inflation and the reheat temperature. 
The explicit breaking of the PQ symmetry (other than the QCD effect) is tiny at present because it is effective only at the Planck scale VEV, at which the PQ-symmetry breaking field is kicked in the phase direction in the early universe. Moreover, in the minimal setup, the PQ-symmetry breaking terms are proportional to the inflaton energy, which also vanishes at present. 
Therefore, the required explicit breaking does not worsen the axion quality problem with regard to solving the strong CP problem.

Typically, parametric resonance~\cite{Kofman:1994rk,Kofman:1997yn} may occur during the radial oscillations due to significant mixing between the radial and angular modes of the PQ breaking field~\cite{Ballesteros:2016euj,Ballesteros:2016xej,Ema:2017krp,Co:2017mop}. 
This effect can be understood by noting that the PQ breaking field has a tachyonic mass at the origin of the potential, where the fluctuations of the PQ breaking field are excessively amplified. 
This poses threats to the models because, if the PQ symmetry is non-thermally restored by parametric resonance, cosmic strings and domain walls will form and are cosmologically stable for domain wall numbers greater than unity~\cite{Co:2020dya,Co:2022qpr}. The axions created during parametric resonance may also become too warm to be dark matter~\cite{Co:2017mop,Co:2020dya,Co:2020jtv}. One way to avoid the issues is to thermalize the PQ breaking field so that the elliptical motion becomes circular and therefore cannot trigger parametric resonance.  In the current scenario, the potential is initially dominated by the purely quadratic Hubble-induced terms at the onset of oscillation, and parametric resonance can be delayed due to the lack of self-interactions until the bare mass dominates. This delay may relax the condition to avoid catastrophe because thermalization can now occur at a later time.

We will show how both DM and baryon asymmetry can be explained simultaneously in our scenario. 
We also discuss a supersymmetric (SUSY) model as a UV completion for our scenario in Appendix~\ref{sec:model}. 
In particular, we do not need a strong SUSY breaking effect because the PQ breaking field is kicked by Hubble-induced terms at the end of inflation 
rather than soft SUSY breaking terms.

Here we summarize the advantages of our scenario; 
i) no isocurvature perturbations (i.e., no domain-wall or isocurvature problems),
ii) higher likelihood of avoiding parametric resonance and thus the dangerous topological defects and/or hot axion dark matter
iii) independence of the details of the Planck-suppressed operator and the SUSY sector, especially for high-scale inflation, 
iv) consistency with the axion quality problem, 
v) compatibility with low-energy SUSY breaking models, 
and 
vi) realizability in SUSY (or supergravity) models.

The rest of this paper is organized as follows. 
In Sec.~\ref{sec:general}, 
we explain the basic idea of the flipped Hubble-induced mass for the PQ breaking field 
and solve its dynamics to calculate the energy density and angular velocity of the axion. 
We derive an upper bound on the bare mass for the PQ breaking field by requiring that its energy density does not dominate before it is dissipated into the thermal plasma. 
In Sec.~\ref{sec:main}, 
we show the parameter space in which we can explain the abundance of DM and baryon asymmetry of the Universe. 
In Appendix~\ref{sec:model}, 
we consider a UV complete model based on SUSY. 
Finally, we discuss and conclude in Sec.~\ref{sec:discussions}.

\section{Axion rotations from Hubble-induced masses}
\label{sec:general}

\subsection{Flipped Hubble-induced mass term}

We denote $I$ and $P$ as the inflaton and PQ breaking field, respectively. 
We assume that they couple with each other via the Planck-suppressed operators such as 
\beq
 -\mathcal{L} &\supset& 
 - c_{V1} \frac{V(I)}{3 \Mpl^2} \abs{P}^2
 +  c_{K1} \frac{\abs{\del I}^2}{3 \Mpl^2} \abs{P}^2
 + c_{V2} \frac{V(I)}{3 \Mpl^{2M}} \abs{P}^{2M}
 \nonumber
 \\
  &&\qquad 
  - \lkk c_{V3} \frac{V(I)}{3 \Mpl^{N}} P^N
 +  c_{K2} \frac{\abs{\del I}^2}{3 \Mpl^N} P^N 
 + {\rm (c.c.)} 
 \rkk, 
 \label{eq:fullV}
\eeq
where $M$ and $N$ ($\ge 2$) are integers, 
$c_{V1}, c_{K1}, c_{V2}$ are real parameters, 
and $c_{V3}, c_{K2}$ are complex parameters. This is motivated by supergravity~\cite{Affleck:1984fy,Dine:1995kz} as we will explain in Appendix~\ref{sec:model}. 
Other Planck-suppressed terms which we neglect here do not change our discussion qualitatively. 
We also assume that the low-energy potential for the PQ breaking field is negligible at the Planck scale.

During inflation, the kinetic energy of the inflaton is much smaller than the potential 
\beq
 \abs{\del I}^2 \ll V(I) \simeq  3 H_I^2 \Mpl^2, 
\eeq
where $H_I$ represents the Hubble parameter during inflation. 
Substituting these into \eq{eq:fullV}, we obtain 
the effective potential for $P$ as
\beq
 V(P) &\simeq& - \frac{1}{2} c_{V1} H_I^2 (t) S^2  
 + \frac{c_{V2} H_I^2 (t)}{2^M \Mpl^{2M-2}} S^{2M} 
 \nonumber\\
 &&\qquad - \frac{\abs{c_{V3}} H_I^2 (t)}{2^{N/2-1} \Mpl^{N-2}} S^{N} \cos (N \theta - \delta_{V3})
,
 \label{V-inf}
\eeq
where we rewrite $c_{V3} = \abs{c_{V3}} e^{-i \delta_{V3}}$ and 
\beq
 P = \frac{1}{\sqrt{2}} S e^{i \theta},
\eeq
with the radial mode $S$ we call the saxion.
Hereafter, we take $\delta_{V3} = 0$ without loss of generality by shifting the phase direction $\theta$. 
When $c_i$ are $\mathcal{O}(1)$ and $c_{V1} > 0$, 
the radial and phase directions have masses of order $H_I$ and relax towards the potential minima at 
\beq
 && S \sim \sqrt{2} \lmk \frac{c_{V1}}{c_{V2}} \rmk^{1/(2M-2)} \Mpl
 \label{eq:VEV}
 \\
 && \theta \simeq 0. 
\eeq
As a result, the quantum fluctuations for phase direction are exponentially damped during inflation. 
The isocurvature problem and domain wall problem are therefore absent in this setup.

After inflation, the inflaton $I$ starts to oscillate around its potential minimum. The energy density of the universe is dominated by its oscillation energy until it completely decays into radiation. The Hubble parameter decreases as $H(t) \simeq H_I (a_I/a(t))^{3/2}$, where $a(t)$ is the scale factor and $a_I$ is the scale factor at the end of inflation. 
During the inflaton-oscillation dominated epoch, we expect
\beq
 \frac{\left\langle  V(I) \right\rangle}{3 \Mpl^2} \simeq  \frac{H^2(t)}{2}
 \\
 \frac{\left\langle  \abs{\del I}^2 \right\rangle}{3 \Mpl^2} \simeq  \frac{H^2(t)}{2}
\eeq
after taking the time average denoted by the brackets. 
The effective potential of $P$ is then given by 
\beq
 V(P) &=& \frac{1}{2} \lmk \frac{c_{K1} - c_{V1}}{2} \rmk H^2 (t) S^2  
 + \frac{c_{V2} H^2(t)}{2^{M+1} \Mpl^{2M-2}} S^{2M} 
 \nonumber\\
 &&\qquad - \frac{c_\theta H^2(t)}{2^{N/2} \Mpl^{N-2}} S^{N} \cos \lmk N \lmk \theta - \delta \rmk \rmk
. 
 \label{V-osc}
\eeq
Here, $c_\theta$ and $\delta$ are $\mathcal{O}(1)$ real numbers defined by $c_{V3} + c_{K2} = c_\theta e^{-i \delta}$.

In this paper, we focus on the case with $c_{K1} > c_{V1} > 0$. Then the first term in \eq{V-osc} is positive and the field $P$ starts to oscillate around $P \simeq 0$ after inflation~\cite{Yamada:2015xyr}. 
At the same time, the phase direction is kicked by the last term in \eq{V-osc} because the minimum of the phase direction changes from $\la \theta \ra = 0$ to $\delta$ at the end of inflation. 
The PQ charge is produced via the dynamics. By numerical analyses, we verify that the phase direction does not adiabatically track the minimum as it evolves, and the rotation in the phase direction is indeed generated. 
We find that the initial angular velocity can be as large as $\dot\theta = \mathcal{O}(0.1) H_I$.

Note that the energy density of $P$ starts with being comparable to that of the inflaton but redshifts faster because $\rho_P(t) \simeq H(t) n_{\rm PQ}(t) \propto a^{-9/2}(t)$. 
After reheating completes, the dynamics of the PQ breaking field is non-trivial because the Hubble-induced terms from \eq{eq:fullV} are absent and a thermal potential appears. 
We also need to specify its low-energy potential to discuss its detailed dynamics. 
In the subsequent subsections, we consider its dynamics and check that the PQ breaking field does not dominate.

\subsection{Dynamics of the PQ breaking field}

Now we examine the dynamics of the PQ breaking field after inflation and derive how various quantities evolve. 
For this purpose, we need to specify a low-energy potential for the radial mode. We consider the case where the potential is nearly quadratic as motivated in supersymmetric models with the following two explicit examples, including
\beq
 V_0(P) &=& \frac{1}{2} m_0^2 \abs{P}^2 \lmk \ln \frac{2 \abs{P}^2}{f_a^2} - 1 \rmk , 
\eeq
where the logarithmic correction arises from the soft mass running of the $P$ field, and a two-field model
\beq
W&=&X(P\tilde P - v^2), \hspace{0.5cm} V=m_P^2 \abs{P}^2 + m_{\tilde P}^2 \abs{\tilde P}^2,
\eeq
where $X$ is a field whose $F$-term potential fixes $P$ and $\tilde P$ to a moduli space, which is then lifted by the soft masses $m_P$ and $m_{\tilde P}$.
This low-energy potential becomes important for the dynamics of $P$ at the late stage. 

For the purpose of the PQ field thermalization, we will consider a coupling between the PQ breaking field and heavy vector quarks, which can be identified as the KSVZ quarks in the Kim-Shifman-Vainshtein-Zakharov (KSVZ) model~\cite{Kim:1979if,Shifman:1979if} or need to be added to the Dine-Fischler-Srednicki-Zhitnitsky (DFSZ) model~\cite{Zhitnitsky:1980tq,Dine:1981rt}.

The dynamics of the PQ breaking field can be divided into three regimes by considering which (effective) potential dominates its dynamics.

\ 

\noindent 
{\bf Hubble-induced potential}

After the onset of oscillations, 
the effective potential of $P=S e^{i \theta}/\sqrt{2}$ is given by 
\beq
 V_H (P) \simeq \frac{1}{2} c_H H^2(t) S^2, 
\eeq
where $c_H \equiv (c_{K1} - c_{V1})/2$ is a positive $\mathcal{O}(1)$ parameter. 
The dynamics is similar to the harmonic oscillator with an adiabatically varying frequency. In this case, the comoving ``number density" of the radial oscillation $n_{\rm PQ} a^3(t) \simeq m_{\rm eff}(t) \bar{P}^2(t) a^3(t)$ should be conserved, where $m_{\rm eff} \simeq \sqrt{c_H} H(t)$ is an effective mass for the radial direction. We note that the comoving PQ charge density $\dot{\theta} \bar{P}^2 a^3$ is conserved as well. 
Thus the amplitude $\bar{P}$, energy density $\rho_P$, and time-averaged angular velocity $\dot{\theta}$ of $P$ decreases as 
\beq
 &&\bar{P}(t) \simeq S_{\rm osc} \lmk \frac{a(t)}{a_{\rm I}} \rmk^{-3/4} 
 \label{eq:St}
 \\
 &&\rho_P \simeq H(t) n_{\rm PQ} \simeq H_I^2 S_{\rm osc}^2 \lmk \frac{a(t)}{a_I} \rmk^{-9/2}, 
 \\
 &&\left\langle \dot{\theta} \right\rangle \simeq  H_I \lmk \frac{a(t)}{a_{\rm I}} \rmk^{-3/2}, 
 \label{eq:dottheta0}
\eeq
where $S_{\rm osc}$ is the saxion field value at the onset of oscillations and is expected to be $\mathcal{O}(\Mpl)$. We note that we are interested in the time average $\left\langle \dot\theta \right\rangle$ of $\dot\theta$ for the purposes of baryogenesis. Although the initial angular velocity $\dot\theta \simeq \epsilon H_I$ depends on $\epsilon \lesssim 0.1$ that parametrizes the strength of the kick, the time average value is independent of $\epsilon$~\cite{Co:2020jtv}.

\

\noindent 
{\bf Thermal-log potential}

During the inflaton-oscillation dominated epoch, 
some fraction of the inflatons decays into radiation and the ambient plasma grows in the background. 
Assuming that radiation will be thermalized within one Hubble time,%
\footnote{
This assumption does not always hold, but for a relatively high reheat temperature, it is a good approximation~\cite{Harigaya:2013vwa,Mukaida:2015ria,Mukaida:2022bbo}.
}
its temperature is given by (see e.g., Ref.~\cite{Harigaya:2014waa})
\beq
 T \simeq \lmk \frac{36 H(T) \Gamma_I \Mpl^2}{g_* \pi^2} \rmk^{1/4} \propto a^{-3/8}, 
\eeq
where $\Gamma_I$ is the inflaton decay rate. 
The reheat temperature $T_{\rm RH}$ is defined by the temperature at the end of reheating and is given by 
\beq
 T_{\rm RH} \simeq \lmk \frac{90}{g_* \pi^2} \rmk^{1/4} \sqrt{\Gamma_I \Mpl}. 
\eeq

In this background, the PQ breaking field acquires an effective potential via the thermal effect. 
Since the amplitude of $P$ is much larger than the temperature of the plasma, 
it acquires the so-called thermal-log potential via two-loop effects~\cite{Anisimov:2000wx,Fujii:2001zr}. 
This can be understood by noting that the renormalization-group running of gauge coupling constants changes as the mass of heavy fields changes. Since the effective mass for fields that couple to the PQ breaking field depends on $\abs{S}$, the running of the gauge coupling also depends on $\abs{S}$ logarithically. As the energy density of the thermal plasma has a correction of order $\alpha T^4$ from the one-loop effect, these effects generically result in 
\beq
 V_T(P) \simeq \alpha^2 T^4 \log (\abs{S}^2 / T^2), 
\eeq
where $T$ is the temperature and $\alpha$ represent a fine-structure constant for the SM gauge interaction, which we assume is $\simeq 1/27$ at high temperatures. 
This term comes from the change of the renormalization group running of the gauge coupling by the VEV-dependent mass of heavy quarks.

For the case with
\begin{equation}
\label{eq:therm_log_condition}
    H_{\rm RH} \lesssim \frac{90}{\pi^2 g_*} \alpha^2 H_I \lmk\frac{\Mpl}{S_{\rm osc}}\rmk^2 ,
\end{equation}
the thermal-log term can dominate the potential before reheating completes. 
Denoting the Hubble parameter at this threshold as $H_{V_T}$, 
we obtain 
\beq
 H_{V_T} \simeq \alpha  \sqrt{ H_{\rm RH} H_I \frac{90}{\pi^2 g_*}} \, \lmk \frac{\Mpl}{S_{\rm osc}} \rmk. 
\eeq
Noting that the effective mass of the PQ breaking field is now given by $m_{\rm eff} \sim \alpha T^2(t) / \abs{S}$ and $T \propto a^{-3/8}$ and matching the values of $\bar{P}$ and $\left< \dot{\theta} \right>$ with Eqs.~(\ref{eq:St}) and (\ref{eq:dottheta0}) at $H(t) = H_{V_T}$, we can write
the amplitude, energy density, and time-averaged angular velocity of $P$ as 
\beq
 &&\bar{P}(t) \simeq S_{\rm osc} \lmk \frac{H_{V_T}}{H_I} \rmk^{1/2} \lmk \frac{H(t)}{H_{V_T}} \rmk^{3/2}
 \label{eq:S1}
 \\
 &&\rho_P \simeq \alpha^2 T^4(t)  \propto a^{-3/2}(t), 
 \\
 &&\left\langle \dot{\theta} \right\rangle \simeq H_{V_T} \lmk \frac{H(t)}{H_{V_T}} \rmk^{-1},
 \label{eq:dottheta1}
\eeq
for $H_{\rm RH} < H(t) < H_{V_T}$. 
In particular, $\left\langle \dot{\theta} \right\rangle$ grows in this regime and reaches $\epsilon \alpha^2 H_I$. 

For the case opposite to \eq{eq:therm_log_condition}, the thermal-log potential remains smaller than the Hubble-induced potential both before and after reheating. The Hubble-induced mass is absent after inflationary reheating completes. If $m_0 < H_{\rm RH}$, the saxion becomes effectively massless after reheating. Without a sufficient curvature in the radial direction to provide the centripetal force and support the rotation, the saxion field value increases slightly and the rotational motion quickly enters the slow-roll regime by the overdamping Hubble friction. When the Hubble parameter later decreases to the saxion vacuum mass, the PQ field starts to rotate again. Despite the slow-roll regime, the PQ charge is conserved throughout the evolution since the potential is PQ conserving at this radius. Unlike the standard scenario of kinetic misalignment with a Planck-scale initial VEV but no flipped Hubble-induced mass, this scenario is still free of isocurvature perturbations due to post-inflationary relaxation of the angular mode to a local minimum. 
For the purposes of baryogenesis discussed in Sec.~\ref{sec:main}, we will consider the scenario where the thermal mass never dominates---namely, \eq{eq:therm_log_condition} is violated---but the bare mass term $m_0$ dominates over the Hubble-induced mass during reheating, $m_0 > H_{\rm RH}$. The evolution of the rotation smoothly transitions when the quadratic potential dominates, and $\dot\theta$ becomes a constant value $m_0$ until the rotation settles to the minimum of the radial potential.

In the remainder of this section, we focus on the case with \eq{eq:therm_log_condition}. After reheating, the temperature decreases as $T(t) \propto a^{-1}(t)$ in the radiation-dominated era. 
Again, noting that the effective mass of the PQ breaking field is still given by 
$m_{\rm eff} \sim \alpha T^2(t) / \abs{S}$ and matching the values of $\bar{P}$ and $\left< \dot{\theta} \right>$ with Eqs.~(\ref{eq:S1}) and (\ref{eq:dottheta1}) at $H(t) = H_{\rm RH}$, we find
the amplitude, energy density, and time-averaged angular velocity of $P$ given by 
\beq
 &&\bar{P}(t) \simeq 
   \frac{\Mpl}{\alpha} \sqrt{\frac{\pi^2 g_*}{90}}
  \lmk \frac{S_{\rm osc}}{\Mpl} \rmk^2
 \lmk \frac{H_{\rm RH}}{H_I} \rmk  
 \, \lmk \frac{a(t)}{a(t_{\rm RH})} \rmk^{-1}
 \\
 &&\rho_P \simeq \alpha^2 T^4(t)  \propto a^{-4}(t), 
 \\
 &&\left\langle \dot{\theta} \right\rangle \simeq  \lmk \frac{90}{\pi^2 g_*} \rmk \alpha^2  H_I \, \lmk \frac{\Mpl}{S_{\rm osc}} \rmk^2
 \lmk \frac{a(t)}{a (t_{\rm RH})} \rmk^{-1}.
 \label{eq:dottheta2}
\eeq

\ 

\noindent 
{\bf Low-energy potential}

As the temperature decreases, 
the bare mass of $P$ eventually dominates the thermal-log potential. 
The temperature at which the bare mass term comes to dominate is given by 
\beq
T_* 
\simeq \frac{\pi^2 g_* m_0}{ 90 \alpha^2} \lmk \frac{T_{\rm RH}}{H_I} \rmk 
\lmk \frac{S_{\rm osc}}{\Mpl} \rmk^2.
\eeq
Then 
we obtain 
\beq
 &&\bar{P} (t) \simeq \alpha \frac{T_*^2}{m_0} \lmk \frac{a(t)}{a (t_*)} \rmk^{-3/2} 
 \nonumber\\
 &&\rho_P (t) \simeq \alpha^2 T_*^4 \lmk \frac{a(t)}{a (t_*)} \rmk^{-3}  
  \nonumber\\
 &&\left\langle \dot{\theta} \right\rangle \simeq m_0 
\eeq
for $T < T_*$. 
After the amplitude $S(t)$ becomes as low as $f_a$, 
the radial direction settles to the potential minimum. 
We denote this time scale as $t_S$. 
The angular velocity is then given by 
\beq
 \left\langle \dot{\theta} \right\rangle \simeq 
m_0
\lmk \frac{a(t)}{a (t_S)} \rmk^{-3}. 
\eeq

\subsection{Thermalization of the radial direction}

In the previous subsection, we assume that the radial direction is not thermalized by the scattering with the thermal plasma. When the vacuum potential dominates, the energy density decreases slower than radiation, so that the radial direction eventually dominates the energy density unless the energy density is depleted sufficiently early, e.g., via dissipation and thermalization. 
If we omit its dissipation, the temperature at which $\rho_P$ comes to dominate is given by 
\beq
T_{\rm dom} = \frac{4 m_0 Y_{\rm PQ}}{3 \epsilon},
\eeq
where the PQ charge yield, defined as the charge-to-entropy density ratio $Y_{\rm PQ} \equiv n_{\rm PQ} /s$, is redshift invariant in the absence of entropy production.

Now we will check if the radial direction thermalizes before its energy density dominates. The dissipation rate for the PQ breaking field is given by 
\beq
 \Gamma_{\rm th} = b \frac{T^3}{\bar{P}^2}, 
\eeq
where $10^{-5} \lesssim b \lesssim 0.1$ is a constant depending on the couplings with gluons or heavy quarks~\cite{Mukaida:2012qn,Mukaida:2012bz}. 
In order for the PQ field not to dominate the energy density before the thermalization, we require $\Gamma_{\rm th} > H$ at $T=T_{\rm dom}$. 
Comparing this with the Hubble expansion rate and using $Y_{\rm PQ} = \epsilon m_0 \bar{P}^2 / (2\pi^2 g_* T^3/45)$, we obtain an upper bound on the bare mass as
\beq
\label{eq:m_0}
m_0 \lesssim 3 \times 10^8 \GeV ~\epsilon^3 
\lmk \frac{b}{0.1} \rmk 
\lmk \frac{228.75}{g_*} \rmk^{3/2} 
\lmk \frac{10^2}{Y_{\rm PQ}} \rmk^3 .
\label{eq:constraint}
\eeq
When this is satisfied, there is no entropy production after reheating, which we assume hereafter. 
In the following section, we will discuss the amount of the charge yield $Y_{\rm PQ}$ that is generated by the Hubble-induced masses as well as the amounts needed to generate the dark matter abundance and the baryon asymmetry.

\section{Kinetic misalignment and (lepto-)axiogenesis}
\label{sec:main}

\subsection{Dark matter density}

The equation of motion for the PQ charge density is given by 
\beq
 \frac{1}{a^3(t)} \frac{\del}{\del t} 
 \lmk a^3(t) n_{\rm PQ}(t) \rmk 
 = -  \frac{\del V}{\del \theta} \, .
 \label{eq:nbl}
\eeq
We obtain 
\beq
 \frac{a^3 (t)}{a^3 (t_{\rm osc})} n_{\rm PQ} (t) \equiv \epsilon H_{\rm osc} S^2_{\rm osc}, 
\eeq
with $\epsilon \sim \delta \sim 1$ being an (inverse) ellipticity parameter. 
Here we denote $H_{\rm osc}$ and $S_{\rm osc}$ as the Hubble parameter and the amplitude of oscillations at the onset of oscillation. In our scenario, $S_{\rm osc} \sim \Mpl$ from \eq{eq:VEV} and $H_{\rm osc} \sim H_I$ because the PQ breaking field starts to oscillate at the end of inflation.

\begin{figure}
    \centering
\includegraphics[width=1.0\linewidth]{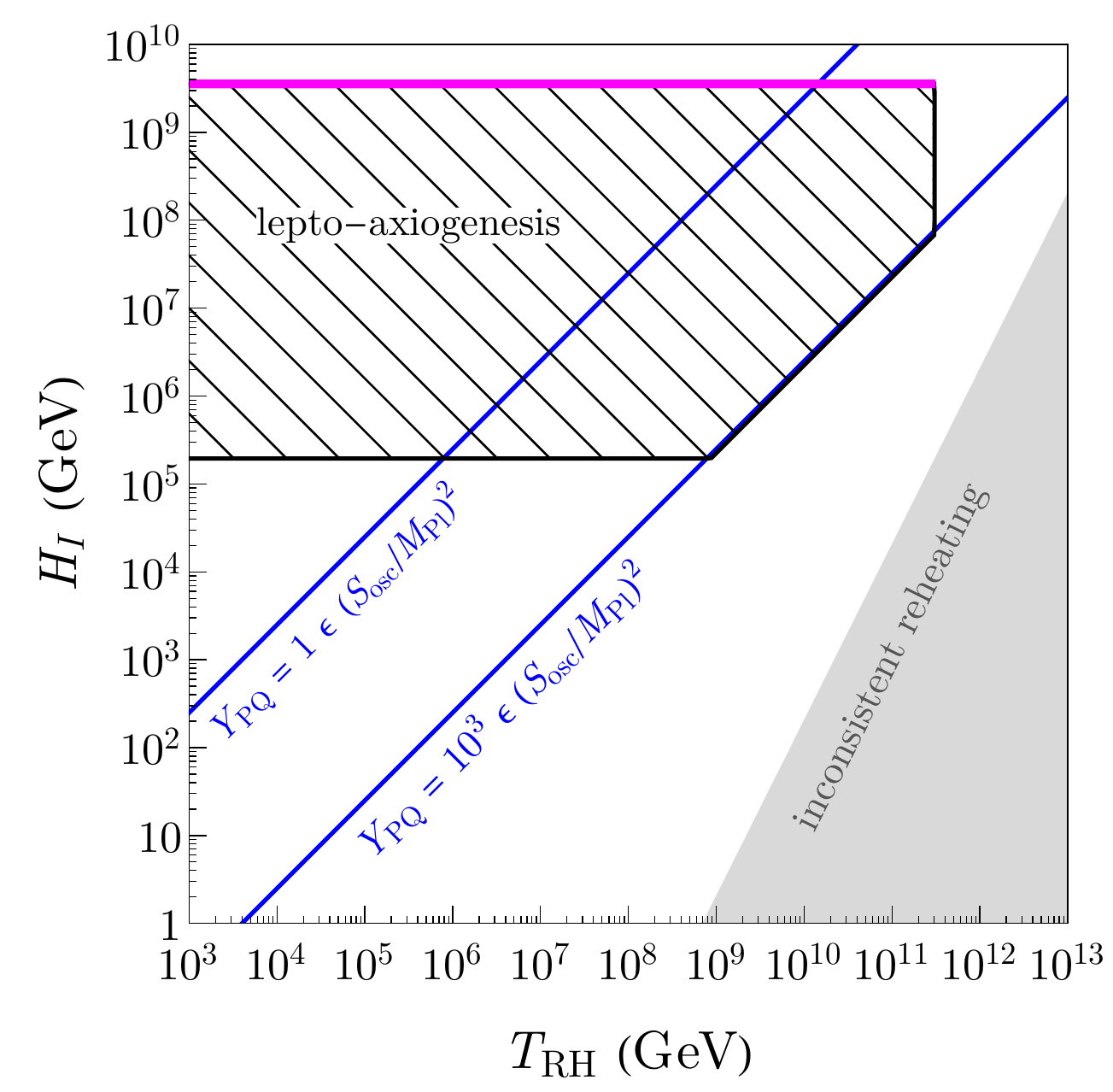}
\includegraphics[width=1.0\linewidth]{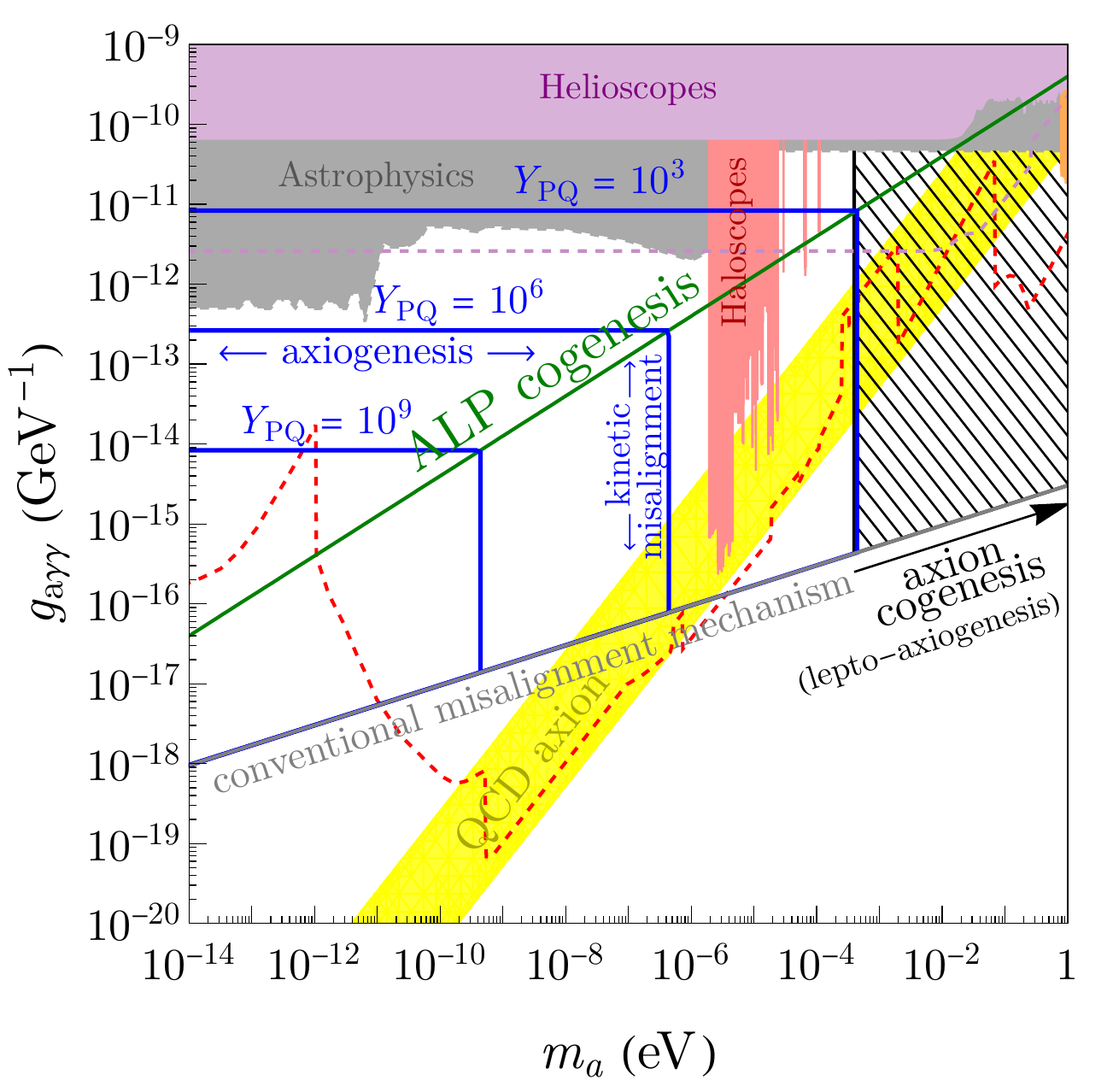}
    \caption{{\bf Top:} Contours of $Y_{\rm PQ}$ generated by the Hubble-induced potential are shown by the blue lines (see \eq{eq:YPQ}). In the black hatched region and along the magenta line, the observed baryon asymmetry can be explained by lepto-axiogenesis. The vacuum potential (the thermal-log potential) dominates at $T_{\rm RH}$ in the black hatched region (along the magenta line). The gray region is self-inconsistent because $H_{\rm RH} > H_I$. 
    {\bf Bottom:} Contours of $Y_{\rm PQ}$, shown in thick blue lines, to explain the baryon asymmetry from axiogenesis (horizontal segments, see \eq{eq:YBaxiogenesis}) or the DM abundance from kinetic misalignment (vertical segments, see \eq{eq:rho_a_s}). 
    With DM explained by kinetic misalignment, the baryon asymmetry is simultaneously explained along the green line by axiogenesis and in the black hatched region by lepto-axiogenesis.  The yellow band is the parameter space motivated by the QCD axion. The other shaded regions are excluded by current experiments, while the colored dashed contours show the future sensitivities.
    }    
    \label{fig:YPQ}
\end{figure}

After the onset, the oscillation amplitude decreases due to the expansion of the universe. 
From \eq{eq:St}, the oscillation amplitude decreases as $\propto a^{-3/4}$. 
The last term in \eq{V-osc} is then negligible soon after the oscillation begins and the comoving number density of PQ charge becomes approximately conserved, so that $n_{\rm PQ}/\rho_I$ is constant during reheating. If there is no entropy production after inflation, e.g., by the thermalization of the PQ breaking field, the yield $Y_{\rm PQ}$ becomes a constant and is given by
\begin{eqnarray}
Y_{\rm PQ} &\equiv& \frac{n_{\rm PQ}}{s} 
= \left. \frac{n_{\rm PQ}}{\rho_I} \right\vert_{T=T_{\rm osc}} \times \left. \frac{\rho_I}{s} \right|_{T=T_{\rm RH}}
\nonumber\\
&=& \left. \frac{3T_{\rm RH}}{4} \frac{n_{\rm PQ}}{3 H^2 \Mpl^2} \right\vert_{T = T_{\rm osc}} = \epsilon \frac{T_{\rm RH}}{4 H_{\rm osc}} \frac{ S_{\rm osc}^2}{ \Mpl^2}
\nonumber \\[1ex]
& = & 250 ~\epsilon \lmk \frac{T_{\rm RH}}{10^{12} \GeV} \rmk \lmk \frac{10^{9} \GeV}{H_I} \rmk \lmk \frac{S_{\rm osc}}{\Mpl} \rmk^2 \,.
\label{eq:YPQ}
\end{eqnarray}
The top panel of Fig.~\ref{fig:YPQ} shows the contours of $Y_{\rm PQ}$ in the $H_I \,\text{-}\,T_{\rm RH}$ plane. 
The gray shaded region is excluded because the reheat temperature should satisfy $H_{\rm RH} \le H_I$.

The axion energy density over the entropy density is therefore given by kinetic misalignment~\cite{Co:2019jts} and \eq{eq:YPQ} as
\begin{eqnarray}
\label{eq:rho_a_s} 
\frac{\rho_{a}}{s} 
&\simeq& m_a Y_{\rm PQ} \\
&\simeq & 0.5 \eV 
\lmk \frac{\epsilon}{0.1} \rmk
\lmk \frac{m_a}{20 ~{\rm meV}} \rmk 
\lmk \frac{T_{\rm RH}}{10^{12} \GeV} \rmk 
\lmk \frac{10^9 \GeV}{H_I} \rmk 
\lmk \frac{S_{\rm osc}}{\Mpl} \rmk^2,
\nonumber 
\end{eqnarray}
where the observed DM abundance is $\rho_{\rm DM} / s \simeq 0.44 \eV$.
This is one of the main results of this paper. 
In particular, 
the result is independent of parameters for low-energy potential of the PQ breaking field. 
Also, the reheat temperature can be much higher than the original scenario of the kinetic misalignment mechanism even if the PQ breaking field is as large as the Planck scale during inflation. 

Note that the axion energy density depends on the axion's initial phase via the parameter $\epsilon$. If the axion acquires quantum fluctuations during inflation, it results in isocurvature perturbations that are strongly constrained by Cosmic Microwave Background observations. However, in our scenario with the axion as heavy as the Hubble parameter during inflation, quantum fluctuations are damped as the axion starts to oscillate around the minimum during inflation.

The vertical blue lines in the bottom panel of Fig.~\ref{fig:YPQ} show the required values of $Y_{\rm PQ}$ that can explain the observed DM abundance. The vertical segments stop at the gray line because the contribution to axion dark matter from kinetic misalignment is necessarily subdominant to conventional misalignment below the gray line, which we compute by assuming that the initial misalignment angle is unity ($\theta_i = 1$).
The yellow band represents the QCD axion window, where the mass and decay constant are related to each other, but the photon coupling $g_{a\gamma\gamma}$ is model dependent. Hereafter, we assume $g_{a\gamma\gamma} = \alpha_{\rm EM} / (2 \pi f_a) \simeq 10^{-3}/f_a$.
The other shaded regions are excluded by current experiments, while the dashed regions are future sensitivity curves for ongoing and proposed axion experiments. For both shadings and curves, the helioscopes are indicated by purple, the haloscopes are indicated by red, and astrophysical searches are indicated by gray. For helioscopes, the leading constraint is set by CAST~\cite{CAST:2017uph}, and the leading projection is by IAXO+~\cite{IAXO:2019mpb}. For helioscopes, the leading constraints include ADMX~\cite{ADMX:2009iij,ADMX:2018gho,ADMX:2019uok,ADMX:2021nhd}, RBF+UF~\cite{DePanfilis:1987dk,Hagmann:1990tj}, CAPP~\cite{Lee:2020cfj,CAPP:2020utb,Jeong:2020cwz,Lee:2022mnc,Kim:2022hmg,Yang:2023yry}, ORGAN~\cite{Quiskamp:2022pks,Quiskamp:2023ehr}, HAYSTAC~\cite{HAYSTAC:2018rwy,HAYSTAC:2020kwv,HAYSTAC:2023cam}, and QUAX~\cite{Alesini:2019ajt,Alesini:2020vny,Alesini:2022lnp}, whereas the leading projections include DM-Radio~\cite{DMRadio:2022pkf}, ADMX~\cite{Stern:2016bbw}, ALPHA~\cite{Lawson:2019brd}, CADEx~\cite{Aja:2022csb}, BREAD~\cite{BREAD:2021tpx}, and LAMPOST~\cite{Baryakhtar:2018doz}. For astrophysics, the leading constraints include globular clusters~\cite{Ayala:2014pea,Dolan:2022kul}, Chandra~\cite{Reynes:2021bpe,Reynolds:2019uqt}, magnetic white dwarf~\cite{Dessert:2022yqq}, and pulsars~\cite{Noordhuis:2022ljw}, These curves are taken from Ref.~\cite{AxionLimits}.

\subsection{Baryon asymmetry from axiogenesis}

The nonzero angular velocity can also lead to the generation of baryon asymmetry at the electroweak phase transition. This is called axiogenesis~\cite{Co:2019wyp}. 
The baryon asymmetry from axiogenesis is given by
\begin{align}
\label{eq:YBaxiogenesis}
    Y_B =& \frac{c_B Y_{\rm PQ} T_{\rm ws}^2}{f_a^2} \\
    \simeq& 10^{-10} 
    \lmk \frac{\epsilon}{0.1} \rmk
    \lmk \frac{c_B}{0.1} \rmk 
    \lmk \frac{2 \times 10^7 \GeV}{f_a} \rmk^2 \nonumber\\
    & \hspace{0.75cm}  \times 
    \lmk \frac{T_{\rm RH}}{10^{12} \GeV} \rmk 
    \lmk \frac{10^9 \GeV}{H_I} \rmk 
    \lmk \frac{S_{\rm osc}}{\Mpl} \rmk^2,
\end{align}
where we take the temperature $T_{\rm ws}$, at which the electroweak sphaleron processes go out of equilibrium, to be the value predicted by the Standard Model~\cite{DOnofrio:2014rug}, $T_{\rm ws} = 130 \GeV$. The observed baryon asymmetry is $Y_B^{\rm obs} = 8.7 \times 10^{-11}$~\cite{Planck:2018vyg}.
The horizontal blue lines in the bottom panel of Fig.~\ref{fig:YPQ} represent the value of $g_{a\gamma\gamma}$ that can explain the observed baryon asymmetry for a given $Y_{\rm PQ}$.

We can explain both the DM abundance and the baryon asymmetry at the intersecting point of vertical and horizontal blue lines for a given $Y_{\rm PQ}$ in Fig.~\ref{fig:YPQ}. By changing the value of $Y_{\rm PQ}$, we obtain the green line, on which the cogenesis by kinetic misalignment and axiogenesis works. However, the QCD relation of $m_a \simeq 60 \, {\rm meV} (f_a/10^{8} \GeV)^{-1}$ (indicated by the yellow band in the figure) can be consistent with the green line 
only in the excluded region. 
This cogenesis mechanism does not work for the QCD axion unless $T_{\rm ws}$ is higher than predicted by the SM or $c_B$ is significantly larger~\cite{Co:2019wyp}. 
One may need another mechanism to generate baryon asymmetry for the QCD axion.

\subsection{Baryon asymmetry from lepto-axiogenesis}

Since the reheat temperature can be relatively high in our scenario, 
we can consider lepto-axiogenesis~\cite{Domcke:2020kcp,Co:2020jtv}. 
Let us introduce the $B-L$ breaking Weinberg operator as the source of the Majorana mass term for the left-handed neutrinos to explain the neutrino oscillation data. 
If the universe is dominated by radiation, 
the Weinberg operator is in thermal equilibrium at a temperature higher than
\beq
\label{eq:T_BL}
 T_{B-L} \simeq 8 \times 10^{12} \GeV
 \lmk \frac{g_*}{228.75} \rmk^{1/2}
 \lmk \frac{0.03 \eV^2}{\bar{m}^2} \rmk , 
\eeq
where $\bar{m}^2$ is the sum of active neutrino masses squared (see, e.g., Ref.~\cite{Co:2020jtv}). 
The angular velocity of the PQ breaking field can be understood as a chemical potential for the SM fermions, which biases some asymmetries for the SM charges via the transfer equations. Combining the $B-L$ breaking Weinberg operator and the bias factor from the angular velocity, we can obtain a nonzero $B-L$ asymmetry at a high temperature. 
The resulting $B-L$ asymmetry yield is given by 
\beq
 Y_{B-L} = \frac{n_{B-L}}{s} \simeq \left. \frac{T_{\rm RH}}{T_{B-L}} \frac{c_{B-L} \left\langle \dot{\theta} \right\rangle T^2}{s}
 \right\vert_{T = T_{\rm RH}} ,
\eeq
for $T_{\rm RH} < T_{B-L}$ and 
\beq
 Y_{B-L} = \frac{n_{B-L}}{s} \simeq \left. \frac{c_{B-L} \left\langle \dot{\theta} \right\rangle T^2}{s} \right\vert_{T=T_{B-L}} \lmk \log\lmk\frac{T_{\rm RH}}{T_{B-L}}\rmk + 1 \rmk, 
\eeq
for $T_{\rm RH} > T_{B-L}$, 
where $c_{B-L}$ is an $\mathcal{O}(0.1\,\text{-}\,1)$ numerical factor~\cite{Co:2020jtv,Barnes:2022ren}.

We first consider the case where \eq{eq:therm_log_condition} is satisfied so the thermal mass dominates. From \eq{eq:dottheta2}, we have $\left\langle \dot{\theta} \right\rangle \simeq (90/(\pi^2 g_*)) \alpha^2 H_I (\Mpl/S_{\rm osc})^2$ at $T = T_{\rm RH}$. 
We therefore obtain
\beq
 Y_B &=& \frac{28}{79} \frac{n_{B-L}}{s} 
 \nonumber \\
 &\simeq& \frac{28}{79} \frac{90}{\pi^2 g_*}
 \frac{45}{ 2 \pi^2 g_{*s}} 
 \frac{c_{B-L} c_{Y_B} \alpha^2  H_I}{T_{B-L}}
 \lmk \frac{\Mpl}{S_{\rm osc}} \rmk^2
 \nonumber\\
 &\simeq& 
9 \times 10^{-11} \lmk \frac{c_{B-L}  c_{Y_B} \alpha^2}{0.005} \rmk \lmk \frac{H_I}{10^{9} \GeV} \rmk \lmk \frac{\Mpl}{S_{\rm osc}} \rmk^2 ,
\label{eq:BG}
\eeq
where we define
\beq
 c_{Y_B} \equiv 
 \left\{
 \begin{array}{ll}
 1 &{\rm for} \ T_{\rm RH} < T_{B-L}
 \\ 
 \lmk \frac{T_{B-L}}{T_{\rm RH}} \rmk \lmk \log\lmk\frac{T_{\rm RH}}{T_{B-L}}\rmk + 1 \rmk 
 &{\rm for} \ T_{\rm RH} > T_{B-L}
 \end{array}
 \right. .
\eeq
Here, we take $g_{*s}=g_*= 228.75$ for the effective numbers of degrees of freedom for the entropy density and energy density, respectively. The baryon asymmetry is successfully explained along the magenta contour in the top panel of Fig.~\ref{fig:YPQ} with $\alpha = 1/27$ and $c_{B-L} = 1$.
There is a maximum $T_{\rm RH}$ along this curve set by \eq{eq:therm_log_condition}, which explains why the magenta contour is truncated at large $T_{\rm RH}$. The magenta contour is then independent of $T_{\rm RH}$ in the viable parameter space, and this corresponds to the case $T_{\rm RH} < T_{B-L}$ and $c_{Y_B} = 1$. As a result, the case with $T_{\rm RH} > T_{B-L}$ is never realized here.%
\footnote{More generically, decreasing $c_{B-L}$ allows for larger $T_{\rm RH}$. The contour remains independent of $T_{\rm RH}$, i.e., $T_{\rm RH} < T_{B-L}$, as long as $c_{B-L} \gtrsim 0.0015$. This critical value can be derived by setting the maximum $T_{\rm RH}$ allowed by \eq{eq:therm_log_condition} equal to $T_{B-L}$ in \eq{eq:T_BL} using the value of $H_I$ required by lepto-axiogenesis in \eq{eq:BG}.}
This truncation also gives a maximum $Y_{\rm PQ}$ based on the blue contours or \eq{eq:YPQ}. 
As a result, this lepto-axiogenesis scenario works consistently with the kinetic misalignment mechanism in \eq{eq:rho_a_s} down to the axion mass,
\beq
 m_a \gtrsim 
 30 \meV \ 
 \lmk \frac{10^{-3}}{\sqrt{c_{B-L}}  \epsilon \alpha^2} \rmk 
 \lmk \frac{T_{B-L}}{8 \times 10^{12} \GeV} \rmk^{1/2}. 
\eeq

We now consider the case when \eq{eq:therm_log_condition} is violated.  
In this case, $\dot\theta$ at $T_{\rm RH}$ is given by $\max(H_{\rm RH}, m_0)$ depending on whether the Hubble-induced or the bare mass dominates, and the resultant baryon asymmetry is given by
\beq
 Y_B &\simeq& \frac{28}{79}
 \frac{45}{ 2 \pi^2 g_{*s}} 
 \frac{c_{B-L} \dot\theta(T_{\rm RH})}{T_{B-L}}
 \nonumber\\
 &\simeq& 
9 \times 10^{-11} c_{B-L} \lmk \frac{\dot\theta(T_{\rm RH})}{200 \TeV}\rmk .
\eeq
As a result, with a bare saxion mass $m_0 \simeq 200 \TeV 
/ c_{B-L}$, lepto-axiogenesis can explain the baryon asymmetry in the black hatched region in the top panel of Fig.~\ref{fig:YPQ} for $c_{B-L} = 1$. The lower horizontal boundary is set by the condition that $H_I > m_0$ is necessary to relax the PQ field during inflation and generate $Y_{\rm PQ}$ after inflation using Hubble-induced masses. The sloped boundary is a result of $m_0 \simeq 200 \TeV$ violating the thermalization constraint in \eq{eq:m_0} with $Y_{\rm PQ}$ given in \eq{eq:YPQ}. This boundary also sets the smallest axion mass from \eq{eq:rho_a_s}
\beq
\label{eq:ma_min}
m_a \gtrsim 0.4 \meV ~
\lmk \frac{1}{c_{B-L}^{1/3} \epsilon} \rmk
\lmk \frac{0.1}{b} \rmk^{1/3}
\eeq
to explain the dark matter abundance from kinetic misalignment. Lastly, the vertical boundary arises because $\dot\theta(T_{\rm RH}) \ge H_{\rm RH} > 200 \TeV$ will lead to overproduction of the baryon asymmetry at $T_{\rm RH}$.

This cogenesis scenario is shown by the black hatched region in Fig.~\ref{fig:YPQ}. Given the overlap with the QCD axion band, we refer to this as the axion cogenesis region.
The lower bound on the axion mass set by \eq{eq:ma_min} can be translated to $f_a \lesssim 2 \times 10^{10} \GeV$ for the QCD axion. 
The axion decay constant is arbitrary for axion-like particles as long as the dark matter abundance is still produced by kinetic misalignment, which is true above the gray line.

\section{Discussions and conclusion}
\label{sec:discussions}

We have considered a concrete scenario for the axion kinetic misalignment mechanism and (lepto-)axiogenesis where the PQ breaking field stays at the Planck scale during inflation and starts to oscillate just after the end of inflation. 
The phase of the PQ breaking field becomes massive during inflation, so that isocurvature perturbations are absent. This is crucial for kinetic misalignment and 
(lepto-)axiogenesis to explain dark matter and the baryon asymmetry, since otherwise the isocurvature perturbations efficiently grow after inflation via the roulette-like dynamics of the PQ breaking field and may result in the isocurvature or domain-wall problem. 

Thanks to the purely quadratic Hubble-induced potential, this scenario may help avoid parametric resonance, which would otherwise produce dangerous cosmic strings, domain walls, and/or hot axion dark matter. In particular, thermalization transforms the elliptical rotation of the PQ breaking field into a circular one, with which parametric resonance will not occur. An initial potential that is purely quadratic delays parametric resonance and hence increases the likelihood of thermalizing the field prior to parametric resonance. We leave a detailed analysis about this issue for a future work.

The abundance of axion dark matter as well as baryon asymmetry can be simultaneously explained in our scenario by the QCD axion or axion-like particles. 
One may think that, at such a high energy scale, the dynamics of the PQ breaking field is complicated and the predictability would be lost. 
We have discussed that the prediction of the axion abundance and baryon asymmetry can be nearly independent of the parameters in Planck-suppressed operators as well as SUSY breaking terms. For axiogenesis, this is indeed the case.
For lepto-axiogenesis, this is true if the Hubble-induced mass terms initiate the rotation and the mass from the thermal-log potential dominates at the end of inflationary reheating; this case predicts high-scale inflation. 
These scenarios can be realized in supergravity models. 
With lepto-axiogenesis, the QCD axion can achieve cogenesis with a mass $m_a > \mathcal{O}(0.1) \meV$, which would be tested by future axion helioscope and haloscope experiments.

\section*{Acknowledgments}
The authors thank Keisuke Harigaya for comments on parametric resonance and on the case when the thermal mass is irrelevant. We acknowledge the hospitality at APCTP where this work started. 
The work of R.C. was supported in part by DOE grant DE-SC0011842 at the University of Minnesota.
The work of M.Y. was supported by JSPS KAKENHI Grant Numbers
20H05851 and 23K13092, and by the Leading Initiative for Excellent Young Researchers, MEXT, Japan.

\appendix

\section{Model in supergravity}
\label{sec:model}

In this Appendix, we discuss how our scenario naturally arises in SUSY models by including supergravity effects.

In supergravity, the potential and kinetic terms are determined by superpotential $W(\cphi_i)$ and \Kahler potential $K(\cphi_i,\cphi^*_i)$ such as 
\beq
 V_{\rm SUGRA} = e^{K/\Mpl^2} \lkk \lmk D_i W \rmk K^{i \bar{j}} \lmk D_j W \rmk^* - \frac{3}{\Mpl^2} \abs{W}^2 \rkk, 
 \label{SUGRA potential}
\eeq
where 
$D_i W \equiv W_i + K_i W / \Mpl^2$ and $K^{i \bar{j}} = \lmk K_{i \bar{j}} \rmk^{-1}$, 
and 
\beq
 \mathcal{L}_{\rm kin} = K_{i \bar{j}} \del_\mu \cphi^i \del^\mu \cphi^{* j}. 
 \label{kinetic term}
\eeq
The subscript represents the derivative with respect to the corresponding field (e.g., $W_i \equiv \del W / \del \cphi^i$ and $K_{i \bar{j}} \equiv \del^2 K / \del \cphi^i \del \cphi^{*j}$).

In this paper, we assume 
\beq
 K &=& \abs{I}^2 + \abs{\psi}^2 + \abs{P}^2 +  \frac{c_{V1}'}{\Mpl^2} \abs{I}^2 \abs{P}^2 
 \nonumber\\
 &&\quad - \frac{c_{K1}'}{\Mpl^2} \abs{\psi}^2 \abs{P}^2  
 + \frac{c_{V2}'}{\Mpl^{2M}} \abs{I}^2 \abs{P}^{2M} 
 + K_A, 
\label{Kahler}
\eeq
where $I$ is the field whose $F$-term drives inflation and $\psi$ is the field whose oscillation energy dominates after inflation. 
In general, and even in simple inflation models in supersymmetry such as chaotic inflation~\cite{Kawasaki:2000yn,Kallosh:2010ug} and hybrid inflation models~\cite{Copeland:1994vg,Dvali:1994ms}, 
$I$ and $\psi$ are different fields. 
The last term is given by~\cite{Dine:1995kz,Fujii:2002kr,Fujii:2002aj}
\beq
 K_A = - \frac{c_{V3}'}{\Mpl^{N}} \abs{I}^2 P^N 
 - \frac{c_{K2}'}{\Mpl^N} \abs{\psi}^2 P^N + {\rm c.c.}, 
 \label{KA}
\eeq
The Hubble parameter during inflation is given by $H_I^2 \simeq \abs{W_I}^2 / (3 \Mpl^2)$.

The Hubble-induced mass term comes from, e.g., 
\beq
 V_{\rm SUGRA} &\supset& 
 \exp \lmk \frac{K}{\Mpl^2} \rmk 
 W_I K^{I \bar{I}} W_I^* \\
 &\ni& 
 \abs{W_I}^2 \lmk 1 + (1 - c_{V1}') \frac{\abs{P}^2}{\Mpl^2} \rmk, 
 \label{H-mass during inf}
\eeq
during inflation 
and 
\beq
 \mathcal{L}_{\rm kin} \supset K_{\psi \bar{\psi}} \abs{\dot{\psi}}^2 
 \ni \frac{c_{K1}'}{\Mpl^2} \abs{\dot{\psi}}^2 \abs{P}^2. 
\eeq
after inflation. 
Including all contributions, we obtain the effective mass term for $P$ as 
\beq
 V_H &=& c_H H^2 (t) \abs{P}^2 \\
 c_H 
&=& 
\left\{ 
\bea{ll}
- 3 (c_{V1}' - 1) 
&~~~~\text{during \ inflation} \\
3 \lmk - (1-r ) c_{V1}' + r c_{K1}' +  \frac{1}{2} \rmk 
&~~~~\text{after \ inflation}, \\
\eea
\right. 
\nonumber\\
\eeq
where $r$ ($0 \le r \le 1$) is the fraction of the energy density of $\psi$ to the total energy 
after inflation. 
Depending on the parameters $c_{V1}'$ and $c_{K1}'$, 
we can consider $c_H < 0$ during inflation and $c_H > 0$ after inflation. 
Then the field $P$ starts to oscillate after inflation. 

The \Kahler potential in \eq{KA} leads to PQ-symmetry breaking terms like the ones in Eqs.~(\ref{V-inf}) and~(\ref{V-osc}).%
\footnote{See also Refs.~\cite{Fujii:2002aj,Harigaya:2016hqz} for initiating the Affleck-Dine field using the \Kahler potential, whose advantage is to enhanceing the charge by setting the field value at the Planck scale.}
The resulting dynamics of the PQ breaking field is similar to the one discussed in Sec.~\ref{sec:general}. 
We particularly note that 
the condition (\ref{eq:constraint}) should be satisfied to avoid entropy production from the PQ breaking field. 
Assuming a gravity-mediated SUSY breaking model, 
this bound can be understood as the upper bound on the gravitino mass. 
Then we expect that the gravitino mass can be larger than the PeV scale, in which case the reheat temperature can be very high without the gravitino overproduction problem~\cite{Kawasaki:2004qu,Kawasaki:2008qe}. 

On the other hand, one could avoid the gravitino overproduction problem when the gravitino is as light as the $\eV\,\text{-}\,\KeV$ scale~\cite{Moroi:1993mb}. 
Such a low-scale SUSY breaking model is not favored in the original kinetic misalignment mechanism because a relatively large SUSY breaking term is required to kick the PQ breaking field strongly enough~\cite{Co:2019jts}. 
In some parameter space of our scenario, a very light gravitino is allowed because it is the Hubble-induced terms at the end of inflation that kick the PQ breaking field and support the rotation until the end of reheating.

\bibliography{reference}

\end{document}